# Tunable electronic band structure in WS$_{2(1-x)}$Se$_{2x}$ van der Waals Alloys

Meryem Bouaziz[1], Leonard Schue[2], Noeliarinala Felana Andriambelaza[3,4], Natalia Alyabyeva[2], Jean-Christophe Girard[1], Pavel Dudin[5], Fabian Cadiz[2], Jose Avila[5], Yannick Dappe[1], Cesar Gonzalez[6,7], Julien Chaste[1], Fabrice Oehler[1], Christine Giorgetti[3,4], Fausto Sirotti[2], Abdelkarim Ouerghi[1*]

[1]Université Paris-Saclay, CNRS, Centre de Nanosciences et de Nanotechnologies, 91120, Palaiseau, Paris, France
[2]Laboratoire de Physique de la Matière Condensée, CNRS, Ecole Polytechnique, Institut Polytechnique de Paris, 91120, Palaiseau, France
[3]Laboratoire des Solides Irradiés, CNRS, Ecole Polytechnique, CEA/DRF/IRAMIS, Institut Polytechnique de Paris, F-91128 Palaiseau
[4]European Theoretical Spectroscopy Facility (ETSF)
[5]Synchrotron SOLEIL, L'Orme des Merisiers, Départementale 128, 91190 Saint-Aubin, France
[6]Departamento de Física de Materiales, Universidad Complutense de Madrid, E-28040 Madrid, Spain
[7]Instituto de Magnetismo Aplicado UCM-ADIF, Vía de Servicio A-6, 900, E-28232 Las Rozas de Madrid, Spain

The electronic structure of semiconducting 2D materials such as transition metal dichalcogenides (TMDs) is known to be tunable by its environment, from simple external fields applied with electrical contacts up to complex van der Waals heterostructure assemblies. However, conventional alloying from reference binary TMD compounds to composition-controlled ternary alloys also offers unexplored opportunities. In this work, we use nano-angle resolved photoemission spectroscopy (nano-ARPES) and density functional theory (DFT) calculations to study the structural and electronic properties of different alloy compositions of bulk WS$_{2(1-x)}$Se$_{2x}$. Our results demonstrate the continuous variation of the band structure and the progressive evolution of the valence band splitting at the K points from 420 to 520 meV in bulk WS$_{2(1-x)}$Se$_{2x}$. We also carried out scanning tunneling microscopy (STM) measurements and DFT to understand the possible S or Se substitutions variants in WS$_{2(1-x)}$Se$_{2x}$ alloys, with different local atomic configurations. Our work opens up perspectives for the fine control of the band dispersion in van der Waals materials and demonstrate how the band structure can be tuned in bulk TMDs. The collected information can serve as a reference for future applications.

**Keywords:** 2D materials, WS$_{2(1-x)}$Se$_{2x}$ alloys, electronic band structure, photoemission spectroscopy, density functional theory calculations

*Corresponding Author: abdelkarim.ouerghi@universite-paris-saclay.fr

Two-dimensional (2D) materials have emerged as a rich field of study with new solid-state physical properties and high potential value for applications[1]. In parallel of these efforts, active research into the synergistic combination of these atomically thin systems with other functional nanomaterials has also gained increasing interest[2]. In principle, any pristine 2D semiconductor layers with a well-passivated surface should adhere to any other material through van der Waals interactions, while both preserving their structure and physical properties[3,4]. Such an intriguing feature overcomes the critical limitation of epitaxial lattice matching for the fabrication of complex heterostructures[5]. A particular sub-set of 2D materials is that of transition metal dichalcogenides (TMDs) compounds of form MXY with two different chalcogen elements (M = Mo, W and X≠Y = Se, S, Te)[6,7]. These 2D TMDs are ternary alloys, which electronic and structural properties depend on the chemical composition, or ratio between the two chalcogen elements. With equal proportions of both elements, we can obtain the so-called `Janus' structure, where the sub-layers of two different chalcogens (X and Y) are placed symmetrically on the opposite sides of the metal M[8,9,10]. In addition to the spin orbit coupling at K point of reciprocal lattice for single layer, Janus TMDs are reported to show Rashba spin splitting around the Γ point due to the presence of a vertical polarization field, induced by the breaking of the horizontal mirror symmetry[9]. Note that symmetry breaking can also happen in other 2D systems, notably in vertical heterostructures directly obtained from CVD-grown TMDs without transfer[11], with similar induced polarization along the out of plane axis and associated Rashba effects[12]. The existence and the control of the internal electric field at the heterostructure interface is a powerful building block for the engineering of new spintronic devices[13,14].

The direct synthesis of few layer MXY TMDs such as $MoS_{2(1-x)}Se_{2x}$ and $WS_{2(1-x)}Se_{2x}$ is described rather scarcely in the literature[15,7]. Ernandes et al.[7] suggest that $WS_{2(1-x)}Se_{2x}$ ternary alloys offer a promising route to tune the optical and electronic band gap in few layer TMDs. The random local atomic configurations in ternary materials can also break other symmetries, leading to in-plane polarization, as shown recently by Zribi et al[16] on single layer $WS_{1.4}Se_{0.6}$ alloys. The band structure of this alloy shows a clear valence band structure anisotropy characterized by two paraboloids shifted in one direction of the k-space by a constant in-plane vector. By controlling the spatial distribution of chalcogen atoms, the band gap and the overall band structure can be spatially adjusted in an atomically thin 2D layer[7,17] but the effects of random alloy extend to the local electronic structure, spin texture and details of the electronic properties[8]. It is noteworthy that these ternary TMDs alloys are as stable and mechanically robust as their binary counterparts[18]. Nevertheless, the effect on the chemical composition $x$ in $WS_{2(1-x)}Se_{2x}$ on the electronic properties of the bulk material have not been properly explored.

In this study, we investigated $WS_{2(1-x)}Se_{2x}$ alloys to understand how their electronic structure evolves with the S/Se composition. By combining ARPES, Raman spectroscopy, and STM supported by DFT calculations, we correlated global electronic properties with local atomic arrangements. ARPES reveals that the valence band splitting at the K points varies systematically with Se content, while Raman and STM data confirm structural modifications associated with chalcogen substitution and symmetry reduction. STM imaging further highlights local S/Se rearrangements that directly influence the electronic landscape. Together, these results demonstrate that controlling the local chemical environment enables precise tuning of the band structure. This work establishes a pathway for engineering $WS_{2(1-x)}Se_{2x}$ alloys with tailored optoelectronic properties.

The crystal structure of hexagonal $WS_{2(1-x)}Se_{2x}$ monolayer is shown Figure 1(a). Due to the hexagonal symmetry of the monolayer, the in-plane positions of the top and bottom chalcogen sub-layers are superposed[19,20,21]. In order to probe the chemical and structural properties of $WS_{2(1-x)}Se_{2x}$ single crystal (2D semiconductors), micro-Raman and high-resolution x-ray photoemission spectroscopy investigations were performed on the same sample. The chemical composition x is evaluated via photoluminescence (PL) at room temperature on exfoliated single layer (not shown). We observe the expected optical gap decrease as the fraction x in $WS_{2(1-x)}Se_{2x}$ increases[7]. The structural properties of the alloys are investigated by micro-Raman spectroscopy at room temperature. Figure 1(b) shows the representative Raman spectra recorded on bulk crystals of $WS_2$, $WSe_2$ and $WS_{2(1-x)}Se_{2x}$ alloys of various compositions (x=0.2, 0.3, 0.55 and 0.9). In the reference binary samples, the spectrum is dominated by the $A_{1g}$ (S-W, 420 cm$^{-1}$) and the $E_{2g}$ (S-W, 355 cm$^{-1}$) Raman modes for $WS_2$ while the $A_{1g}$ (Se-W) mode dominates at 250 cm$^{-1}$ in $WSe_2$. The crystal quality can be further assessed from the relatively narrow FWHM (~2 cm$^{-1}$) of the respective modes. Consistent with previous studies[22], several spectral changes were observed on the $WS_{1.6}Se_{0.4}$, $WS_{1.4}Se_{0.6}$ and $WS_{0.9}Se_{1.1}$ alloys, while no modification could be detected for the $WS_{0.2}Se_{1.8}$ sample with respect to the reference $WSe_2$. A decrease of the Raman intensity is observed as the S/Se ratio approaches 50%, up to 2 orders of magnitude in the $WS_{0.9}Se_{1.1}$ sample, coherent with the increasing structural disorder induced by the alloying process. In all cases, the three characteristic Raman modes ($A_{1g}$ S-W, $E_{2g}$ S-W, $A_{1g}$ Se-W) of the parent materials are simultaneously detected revealing the presence of both chalcogen atoms in the lattice. Additionally, all Raman peaks exhibit both a significant spectral shift (~5-10 cm$^{-1}$) and broadening (~2-3 cm$^{-1}$) also attributed to the chemical composition in the alloyed samples.

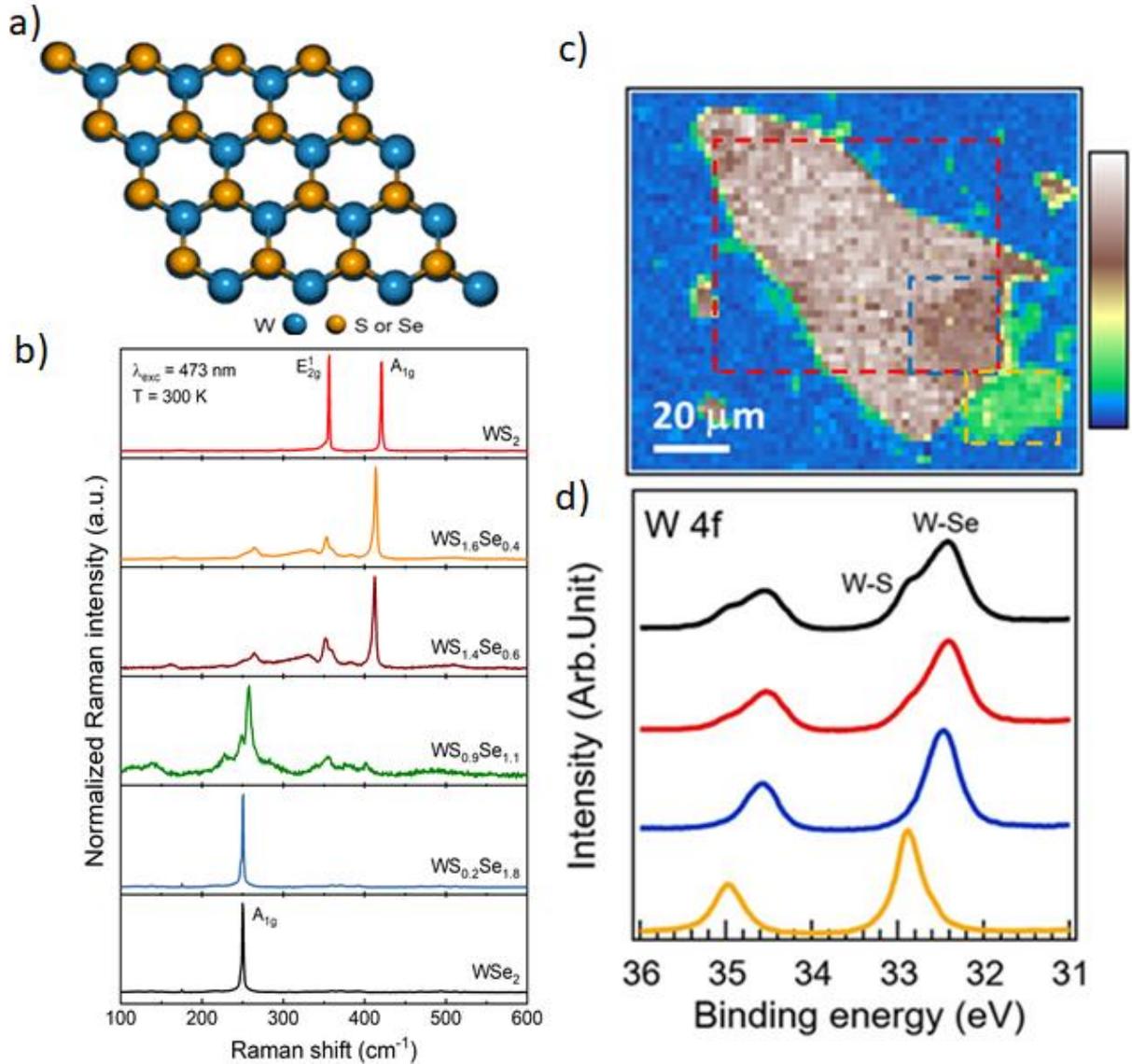

**Figure 1: Optical evidences of the structural of $WS_{2(1-x)}Se_{2x}$ alloys:** a) Atomic structure of $MX_2$ (M =W, X= S, Se), b) Raman spectra recorded at room temperature on bulk crystals of $WS_2$ (top), $WSe_2$ (bottom) and four $WS_{2(1-x)}Se_{2x}$ alloys of various compositions (x=0.2, 0.3, 0.55, 0.9), c) Space-resolved image of the W 4f core level obtained with nano-ARPES experiments on $WS_{0.4}Se_{1.6}$. d) Core level XPS W 4f spectrum of the $WS_{0.4}Se_{1.6}$ measured with hν = 95 eV. The XPS spectra are obtained from the spatial average of the corresponding homogeneous area in b), the color of each curve matches that of the area of origin.

In order to further elucidate the structural and electronic homogeneity of the alloy, we performed spatially resolved nano-ARPES and STM measurements. Nano-ARPES is a powerful technique that enables focusing the X-ray beam to a lateral resolution of about 600 nm, allowing the investigation of the local electronic band structure of $WS_{2(1-x)}Se_{2x}$. An overview XPS spectrum of cleaved bulk $WS_{0.4}Se_{1.6}$ is presented Figure S1, where Se and W peaks are identified. Of particular interest is the presence of two peaks at W core level, related to the two bonds W-S and W-Se. To investigate further, we performed high-resolution space-resolved map of W 4f obtained with nano-ARPES experiments on the sample (Figure 1c) [23]. By integrating the photoemission intensity within a selected energy window around the W 4f, we can get morphological information on the sample. This image shows the inhomogeneity in the composition of the sample by different contrast intensities. To examine the microstructural heterogeneities of $WS_{2(1-x)}Se_{2x}$, different XPS were measured to visualize separate areas containing notably more sulphur or selenium. This image reveals compositional inhomogeneities in the sample, as indicated by the variations in contrast intensity. Figure 1(c) shows that the investigated region exhibits a highly uniform electronic and structural character over an area of approximately 20 µm (blue square), confirming the absence of $WS_2$ or $WSe_2$ phase separation. Figure 1(d) show the representative spectra of the W 4f. The presence of two splitting W 4f core levels is due to the inequivalent structures relative to the bonding with the two chalcogen atoms S and Se. The full surface gives the black curve. The red spectrum is obtained by overlapping the surface in the red box, containing the two contributions with less sulphur. The blue spectrum evidences a zone which is rich in selenium, while the area highlighted in orange gives a spectrum which clearly shows the presence of sulphur. The comparison of the spectra reveals a shift of $W4f_{7/2}$ from 34.4 eV for a selenium-rich area to 32.8 eV with a sulphur-rich composition. The ratio between the area of the S-related component is about 20% of the total one (the sum of the components related to S and Se), which reflects the chemical composition of the alloy. In order to study the electronic properties associated with the modification of the S/Se ratio in the $WS_{2(1-x)}Se_{2x}$ layer, we performed ARPES measurements combined with DFT calculations[24]. Figures 2(a and b) show the ARPES image along the Γ−K−M direction of two compositions ($WS_{1.8}Se_{0.2}$ and $WS_{0.4}Se_{1.6}$). The valence band maximum (VBM) is situated at Γ and is characterized by a parabolic-shaped lobe. Additional sub bands dispersing towards K and M appear and the splitting at the k point is well resolved. The VBM is located at $E-E_F$=-1.10 ± 0.10 eV and -0.9 ± 0.10 eV for $WS_{1.8}Se_{0.2}$ and $WS_{0.4}Se_{1.6}$ respectively. The evolution of the dispersion with binding energy is presented via the constant binding energy cuts in Figure 2(c), which reveal the trigonal symmetry of the dispersion as well as strong intensity variations between the bands in the Brillouin zone of the $WS_{1.8}Se_{0.2}$. The clear definition and sharp spectral signatures of the constant energy contours attest the long-range order of the surface. Moreover, the hexagonal in-plane symmetry of the crystal structure of the films is clearly reflected in their electronic structure, yielding constant energy contours with six-fold symmetry of the $WS_{1.8}Se_{0.2}$ crystal surface. Similarly, the Fermi surface maps of $WS_{0.4}Se_{1.6}$ exhibit the same overall symmetry and band shape as those of the other compounds. The only notable differences are observed in the position of the valence band maximum (VBM) and in the magnitude of the valence band splitting at the K points.

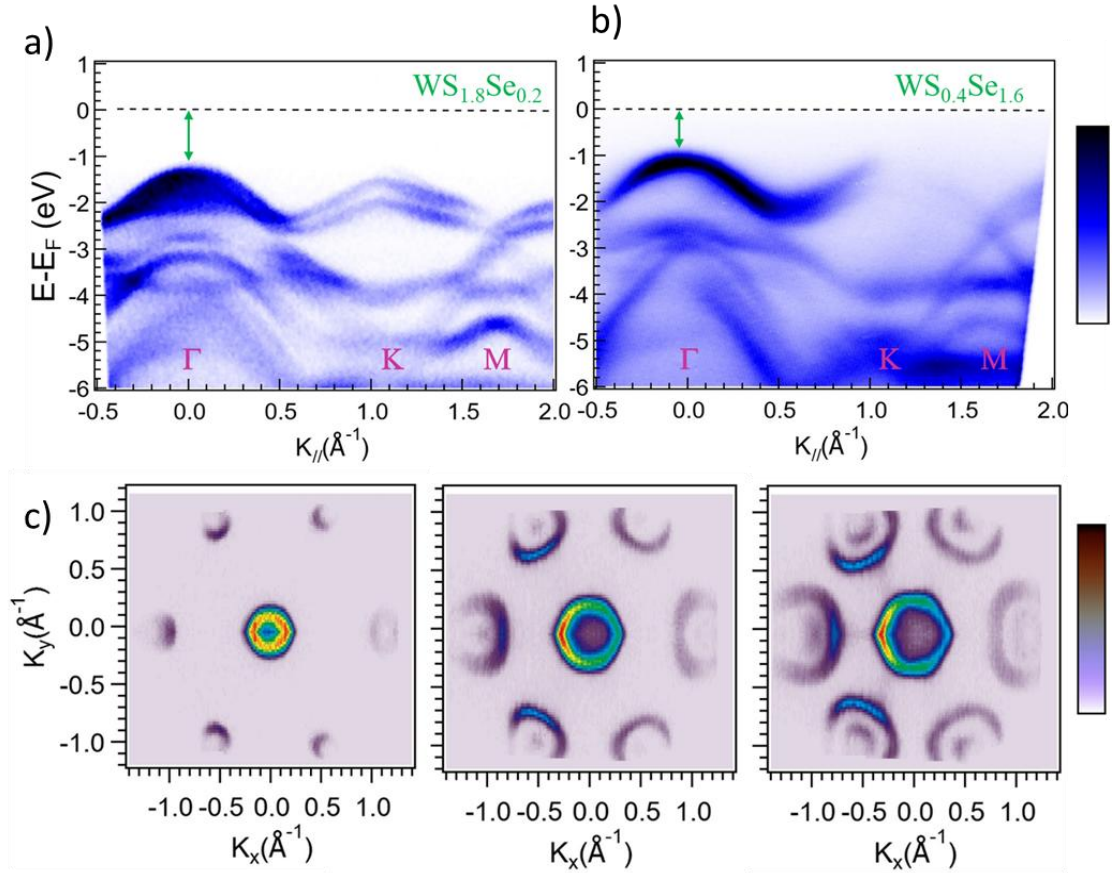

**Figure 2:** ARPES measurements (hν = 95 eV) of the electronic structures along the ΓKM high symmetry directions for a) $WS_{1.8}Se_{0.2}$ and b) $WS_{0.6}Se_{1.4}$. c) Constant energy contours of the $WS_{1.8}Se_{0.2}$ at $E-E_F$ = -1.25 eV, -1.30 eV and -1.35 meV.

Figure 3(a) shows the ARPES image of $WS_{1.8}Se_{0.2}$ along the KΓK direction. The VBM is located at the Γ point and is mostly composed of W $d_{z^2}$ orbitals and $p_z$ orbitals of S and Se. At the K point, it is formed by the d(xy) and d($x^2-y^2$) orbitals of tungsten, hybridized with 3p(x+y) of the chalcogens, as it results from DFT calculations (figure S2). Figure 3(b) displays a comparison of the ARPES data and the band structure calculations for $WS_{1.8}Se_{0.2}$. The band structure was calculated including spin–orbit coupling. Since the compositional disorder extends over large distances, the band structure cannot be reliably obtained from a supercell calculation. Instead, it was derived by averaging the band structures of the two pristine parent compounds, weighted by their respective concentrations. This averaging procedure follows the spirit of Vegard's law[25], assuming that this linear relation can be applied not only to the band gap but also to the eigenvalues of the band structure. To assess the validity of this approach, we compared the so-called averaged band structure for $x = 0.5$ with that obtained from *ab initio* calculations using Quantum ESPRESSO for the Janus WSSe bulk with 2H stacking (Figure S3 and S4). This mean-field approximation is expected to accurately capture the overall trends in the evolution of the band structure[7]. The main features appear to be accurately reproduced by the calculated band structures. The location of the VBM at Γ is consistent with theoretical predictions that bulk $WS_{1.8}Se_{0.2}$ is an indirect-gap semiconductor with the conduction band minimum (CBM) also located at Q point[26]. The effect of chemical composition x in $WS_{2(1-x)}Se_{2x}$ on the band structure near the K point is further investigated by ARPES on samples with x=0, 0.1, 0.7 and 1. Figure 3(c) shows the evolution of the band splitting around K point. We observe that band splitting at the K point increases as more Se is added to the alloy. Contrarily to what is observed for the monolayer, for the bulk materials, the splitting of the two uppermost valence bands at K (K') is not due to spin-orbit coupling (SOC) (Fig S4 left column). The SOC is only responsible for an increase of the splitting, but the value seems to be independent of the chalcogen (Table S1). Moreover, it does not lead to a removal of the spin degeneracy (the criteria to considered that a spin state is polarized is $|S_z| > 0.35$). On notes that for the pristine compounds, the bands are spin-polarized, but degenerated, while for the Janus WSSe with 2H stacking, the bands result to be non spin-polarized. According to the orbital projected band structure (Figure S2), the bands at K results from the hybridization between W $5d(xy+x^2-y^2)$ and S and Se $3p(x+y)$ states, that is only in-plane directions. We conclude that the effect which governs the tuning of the splitting at K, as well as the depth of the bands according to the VBM, is the in-plane overlap between orbitals, which is directly correlated to the in-plane lattice constant (Table S2), itself correlated to the relative concentration of the chalcogens: the larger concentration of Se, the larger the in-plane cell parameter and the larger the splitting. Also, the ARPES data demonstrates that the binding energy of the valence band maximum decreases as a function of increasing selenium content in the bulk alloy, consistent with the decrease in bandgap with increasing selenium content observed in photoluminescence spectroscopy of single layer $WS_{2(1-x)}Se_{2x}$. These nano-ARPES measurements highlight the sensitivity of the valence band to the chemical composition. To further understand how this macroscopic electronic behavior relates to the local atomic environment, we complement these results with STM measurements. STM allows us to directly visualize the atomic-scale arrangement of S and Se atoms and to identify local structural variations that underpin the trends observed in the band structure.

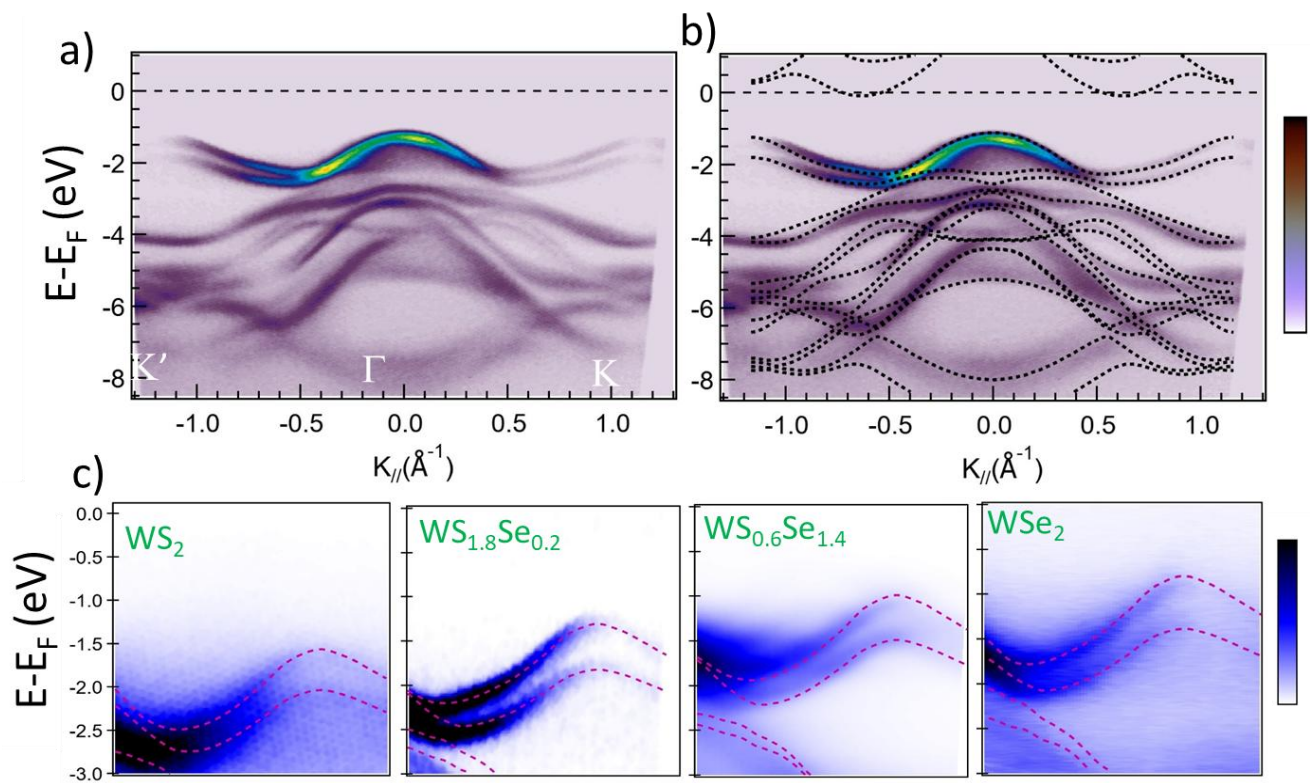

**Figure 3:** a) ARPES map of bulk $WS_{1.8}Se_{0.2}$ along the K'ΓK high-symmetry directions. b) Comparison between ARPES map and DFT calculations. c) Electronic dispersion near the K point for different compositions: $WS_2$, $WS_{1.8}Se_{0.2}$, $WS_{0.6}Se_{1.4}$ and $WSe_2$. Corresponding DFT calculations are superimposed in dotted red.

The same crystals were used for STM experiments cleaving them in vacuum or just before the introduction in the two different microscopes. Figure 4(a) shows the surface of higher S-content $WS_{1.6}Se_{0.4}$ multi-layer, exfoliated over conductive Au surface to ease the STM measurements. Figure 4(b) shows an image from $WS_{0.2}Se_{1.8}$ bulk crystal, prepared in-situ by cleaving the c-plane axis in ultra-high vacuum environment. In both cases we observe the surface of a bulk crystal with atomic resolution. A (1x1) spatial periodicity is observed on both surfaces, consistent with the expected lattice parameters shown Figure 1(a). Native defects, such as selenium vacancies[27] are not visible at this scale (6x6 nm and 8x8 nm). At positions marked (1) and (2) in Figure 4(a) and (b), we clearly image two local configurations with different contrasts (Dark and Bright atoms). The observed contrast difference can be attributed to either the surface's geometric diversity or the electronic state difference at this energy level.

In Figure 4(c) we evidence the presence of the two arrangements using a line profile (from Figure 4(a) black line), and we hypothesize that they correspond to two different configurations with mixed chalcogens: to S over Se (Bright) and Se over S (Dark). The same local arrangements are found in the sample with higher S content in Figure 4(d), around the bright spot (marked in blue) and dark spot (marked in red). Regardless of the $V_{bias}$ applied, S substitutions always appear as depletions with a z-value about 14 pm, corresponding to the difference in ionic radius sizes between the $S^{2-}$ and $Se^{2-}$ ions. As the electronegativity difference between S and Se is very small, the fingerprint of the S atom is essentially topographic in our STM images. Indeed, as shown in a previous study on $MoS_2$[28], at the voltages considered here, the DOS of S (and then Se) is more important than the one of Mo (and here W). Consequently, the Se network always appears brighter and any variation is only due to topographical variations. In order to confirm our hypothesis, we have conducted calculations utilizing the DFT local orbital molecular-dynamics code implemented in Fireball[29]. A 1×1 unit cell of $WS_{2(1-x)}Se_{2x}$ was optimized starting from $WSe_2$ with two structures using single S substitution of the Se in the upper or lower chalcogen sub-layer exclusively. A set of 8×8×1 k-points has been used for both structural optimization and DOS calculations. The results of this study are presented in Figure 4(e), which show the simulated STM height map of $WSe_2$ with upper or lower S substitutions, respectively. These simulated images are constructed with the parameters taken from the experimental conditions of the STM. We observe a strong correlation between the experimental and simulated STM images of $WS_{2(1-x)}Se_{2x}$ which confirm that the substitution of Se by S at the top of the cell (Se bottom) results in a dark atom. Conversely, substitution of Se by S at the bottom of the cell (Se top) results in the formation of a bright atom. Also, We note that the STM images (figure 4(b)) reveal repeated linear arrangements of S- and Se-rich regions extending over several nanometers along specific crystallographic directions. These arrangements are not isolated distortions around single substitution sites but represent a tendency for neighboring substitutional defects to align, suggesting the emergence of nanoscale compositional ordering. The profil line of the STM image (Figure 4(f)) further highlights this directional preference, supporting the interpretation of local, line-shaped ordering rather than random distribution. We therefore attribute these features to a collective structural motif that can influence both the local symmetry and the electronic anisotropy, consistent with the mechanism for in-plane polarization proposed by *Zribi et al.*[16] in the case of single layer WSSe.

Finally, our combined ARPES and STM analysis reveals a direct connection between atomic-scale structure and macroscopic electronic behavior in $WS_{2(1-x)}Se_{2x}$ alloys. The nano-ARPES data reveal that the investigated region exhibits a highly uniform electronic and structural character over an area of approximately 20 µm, confirming the absence of mesoscale phase separation. In contrast, STM investigations expose local, line-shaped arrangements within this otherwise homogeneous region, indicating nanoscale compositional or strain-induced ordering. In this ternary $WS_{2(1-x)}Se_{2x}$ alloy, the partial substitution of one chalcogen species (e.g., sulfur, S) by another of different atomic radius (e.g., selenium, Se) introduces local distortions in the crystal lattice. These substitutions perturb the atomic registry between the chalcogen layers and the intermediate tungsten (W) sublattice, thereby breaking the intrinsic centrosymmetry of the pristine binary TMDs. Our STM results further reveal that the local S- and Se-rich regions form extended, aligned line segments, reflecting a clear in-plane symmetry breaking and giving rise to anisotropic bonding environments at the atomic scale. Such structural anisotropy can induce local dipole moments associated with the polar W–S and W–Se bonds. This microscopic picture supports the mechanism proposed by *Zribi et al.*[16], in which directional ordering of chalcogen atoms and the resulting charge redistribution give rise to a giant in-plane polarization in the case of single layer. Overall, these findings highlight the strong interplay between compositional modulation and lattice distortion, in mixed-chalcogen TMD systems, offering new routes for engineering 2D materials with tunable piezoelectric properties

In summary, we investigate the structural and electronic properties of WSSe van der Waals alloys using combined spatially resolved nano-ARPES and STM measurements. While nano-ARPES reveals mesoscale electronic homogeneity over tens of micrometers, STM uncovers nanoscale, line-shaped compositional arrangements of S- and Se-rich regions. These local substitutions break the in-plane symmetry, inducing anisotropic bonding environments. The nano-ARPES measurements revealed that the binding energy of the valence band maximum decreases with increasing selenium concentration in the bulk alloy. Moreover, the splitting between the two uppermost valence bands at the K point can be continuously tuned from 420 to 520 meV as the Se/S ratio increases. This trend is attributed to the enlargement of the in-plane lattice constant, which modifies the overlap between the in-plane projections of the metal d and chalcogen p orbitals. For all compositions for which ARPES spectra were obtained, the valence band maximum was located at the Γ point.

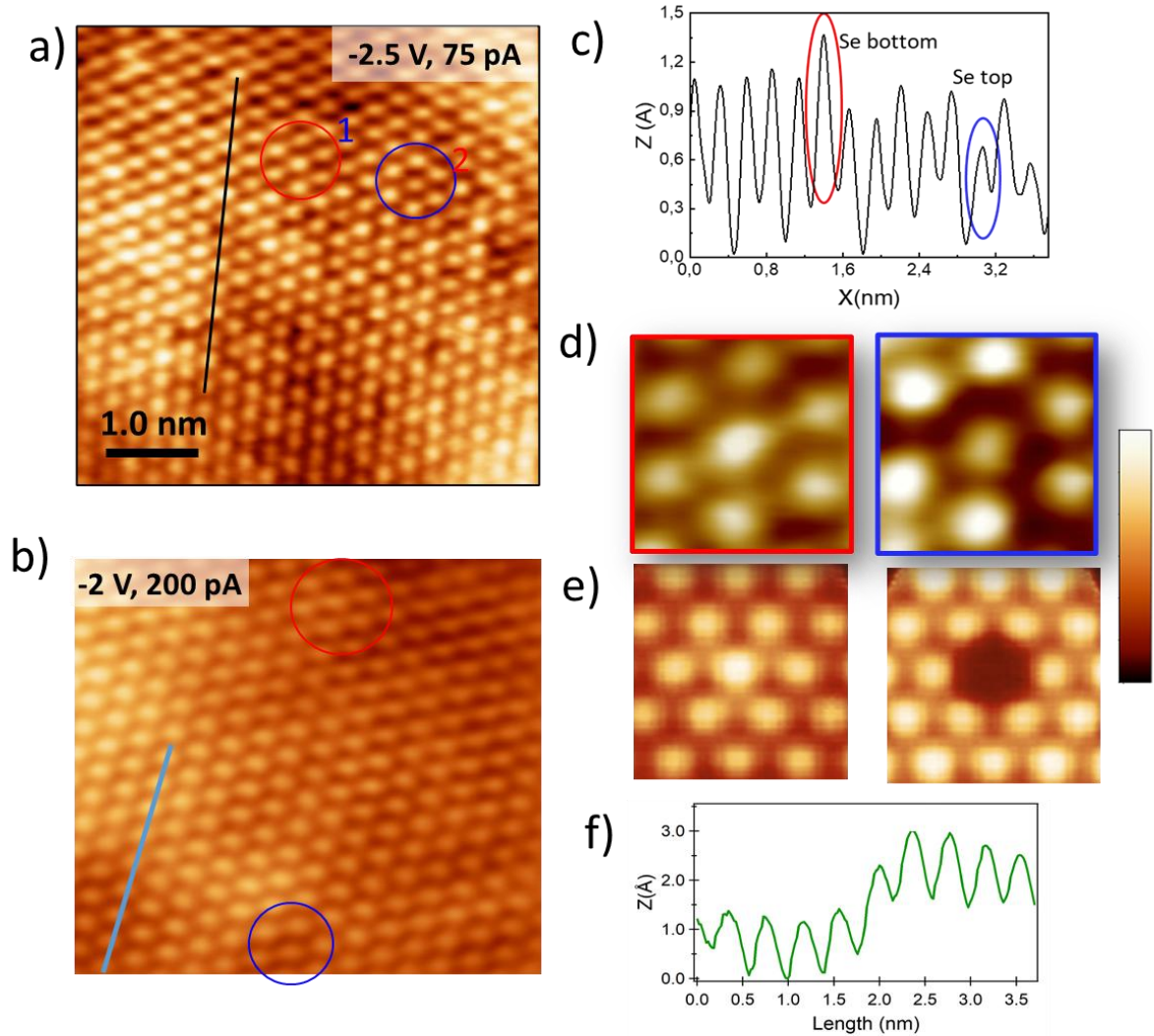

**Figure 4:** a) STM image WS$_{1.6}$Se$_{0.4}$/Au acquired at U = - 2.5 V, I =75 pA. b) STM image of WS$_{0.2}$Se$_{1.8}$ UHV-cleaved surface, acquired at U = - 2.0 V, I =200 pA. c) The line profile from (a) showing the different height assigned to the two chalcogenide elements Se and S. d) local STM images of WS$_{1.6}$Se$_{0.4}$/Au and WS$_{0.2}$Se$_{1.8}$, showing the detailed structure of the two-configuration marked in blue and red in (a). e) Corresponding theoretical simulations. The bright spot represents an S atom substituting a Se atom from the top chalcogen sub-layer, the dark spot is corresponding to a S atom substituting Se in the lower sub-layer. The atomic arrangement is indicated over the figures. Due to a more important DOS with respect to W, any brightness variation of Se in the STM image is only due to topographical reason. f) The line profile from (b) showing the local line-shaped ordering rather than random distribution of the two chalcogenide elements Se and S.

**Materials and methods**

**Experiments:** The Raman spectra were recorded at room temperature in backscattering geometry using a commercial spectrometer (Horiba Jobin-Yvon) and a 473 nm laser excitation (P = 400 µW) with objective x100.

**Photoemission spectroscopy:** Nano-ARPES measurements on WS$_2$$_{(1-x)}$Se$_{2x}$ van der Waals compounds were performed at the ANTARES beamline of the SOLEIL synchrotron facility. The experiments utilized linearly polarized light with a photon energy of 95 eV and a spatial resolution of approximately 600 nm. For the nano-

ARPES measurements, the crystals were prepared by *in situ* cleaving under ultra-high vacuum (UHV) conditions. All measurements were conducted at a base pressure of $3 \times 10^{-10}$ mbar, with the sample temperature maintained at 85 K.

**Computational details:** First-principles calculations for the electronic structure of $WS_{2(1-x)}Se_{2x}$ were performed using density functional theory (DFT) as implemented in the Quantum Espresso (QE) package [1]. We used the projector augmented wave (PAW) pseudopotential [2] to describe the core electron interactions. Exchange-correlation interactions were treated using the generalized gradient approximation (GGA) parametrized by Perdew-Burke-Ernzerhof (PBE) [3]. For the plane-wave expansion of the wave functions, a kinetic energy cutoff of 55 Rydberg was chosen. The structure was fully relaxed until the force on each atom was less than $1.0 \times 10^{-4}$ Ry/Bohr, and the total energy was converged to within $1.0 \times 10^{-5}$ Ry. For both structural relaxation and self-consistent calculations of the band structure, we used a 11 x 11 x 3 k-point mesh to sample the Brillouin zone (BZ). The resulting cell parameters are summarized in table S2. Subsequently, a denser k-point mesh consisting of 153 k-points along the path K-Γ-K in the irreducible Brillouin zone was employed to obtain the band structure. The band structure was calculated including the spin-orbit. Since the disorder occurs on very large distance, the band structure cannot be obtained from a supercell calculation. It was obtained by doing an average of the band structure of the two pristine parents weighted by the respective concentrations.

**Data Availability statements:** data available from the authors upon reasonable request.

**ACKNOWLEDGMENTS:** We acknowledge the financial support by DEEP2D (ANR-22-CE09-0013), 2D-on-Demand (ANR-20-CE09-0026), MixDferro (ANR-21-CE09-0029), Tunne2D (ANR-21-CE24-0030), TyLDE (ANR-23-CE50-0001-01), Optitaste (ANR-21-CE24-0002), ADICT (ANR-22-PEEL-0011) and FastNano (ANR-22-PEXD-0006) projects, as well as the French technological network RENATECH. For this work, access was granted to the HPC/AI resources of IDRIS and TGCC under the allocation No. 0900544 made by GENCI.

**APPENDIX**

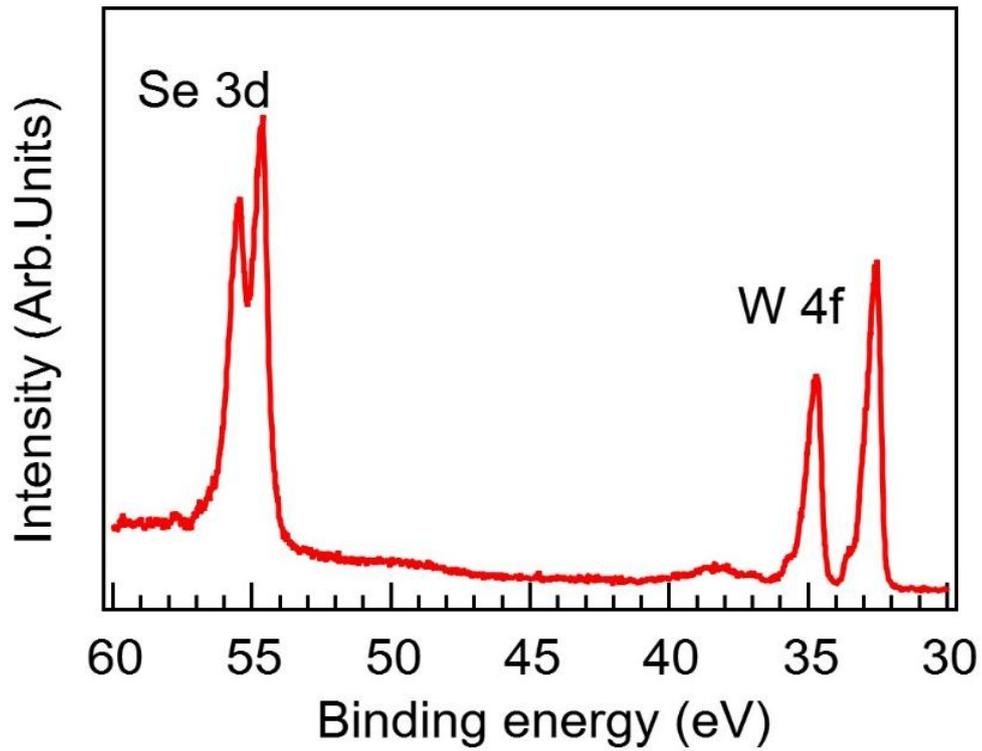

Figure S1: XPS spectrum measured with hν = 95 eV of $WS_{0.4}Se_{1.6}$

To evaluate the effect of the spin-orbit coupling (SOC), we calculated the band structure with and without SOC (Figure S1). We observe that the splitting of the two highest valence bands at K (K') is already present without SOC, which suggests that it is due to the interaction between the two planes of the 2H polytype. The inclusion of SOC does not lead to a removal of degeneracy: the two spin-resolved bands are superposed. The effect of SOC is simply to increase the splitting of the two highest valence bands at K (K') (see Table 1). Interestingly, the increase of the splitting associated with SOC has the same value for both compounds.

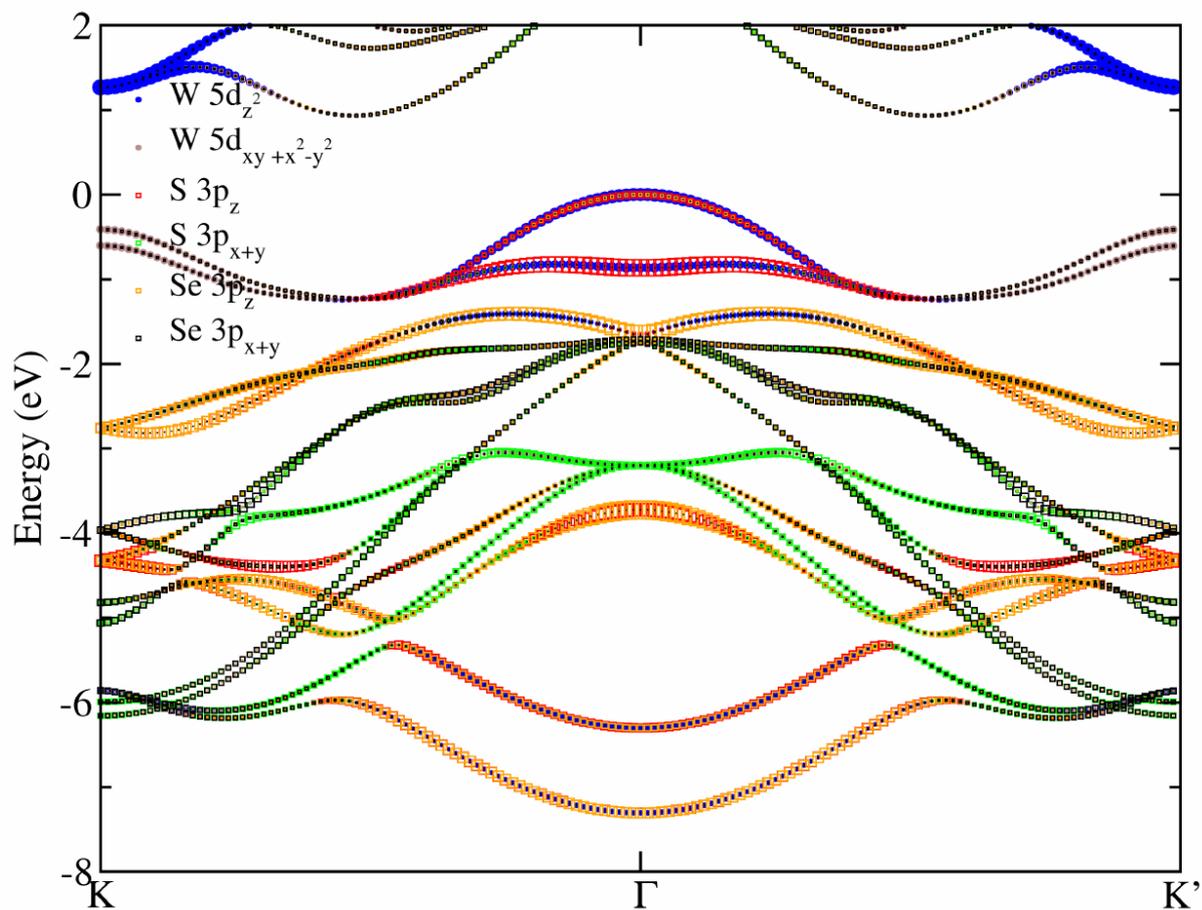
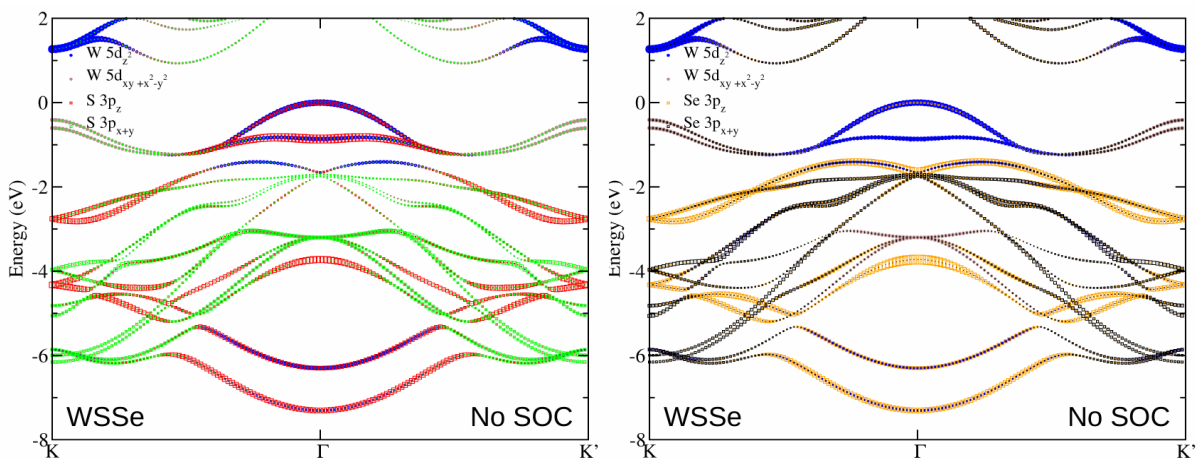

**Figure S2:** (Top) Orbital projected band structure of Janus WSSe. (Bottom) Decomposition of the orbital projected band structure (top) removing bands projected on Se 3p orbitals to evidence the S 3p character (left) and removing bands projected on S 3p orbitals to evidence the Se 3p character (right).

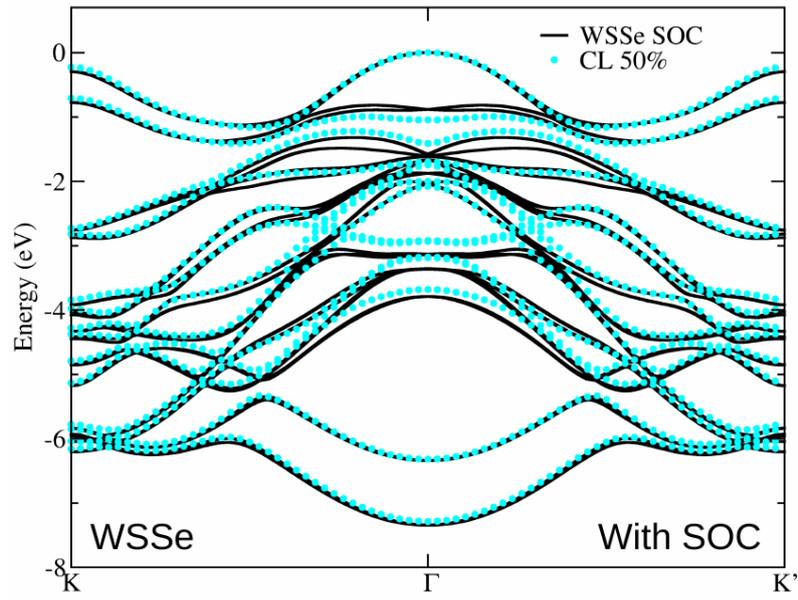

Figure S3: Comparison of the band structure including SOC calculated for the Janus WSSe 2H bulk, compared with the linear combination of the band structure of $WS_2$ and $WSe_2$ with an equal weight.

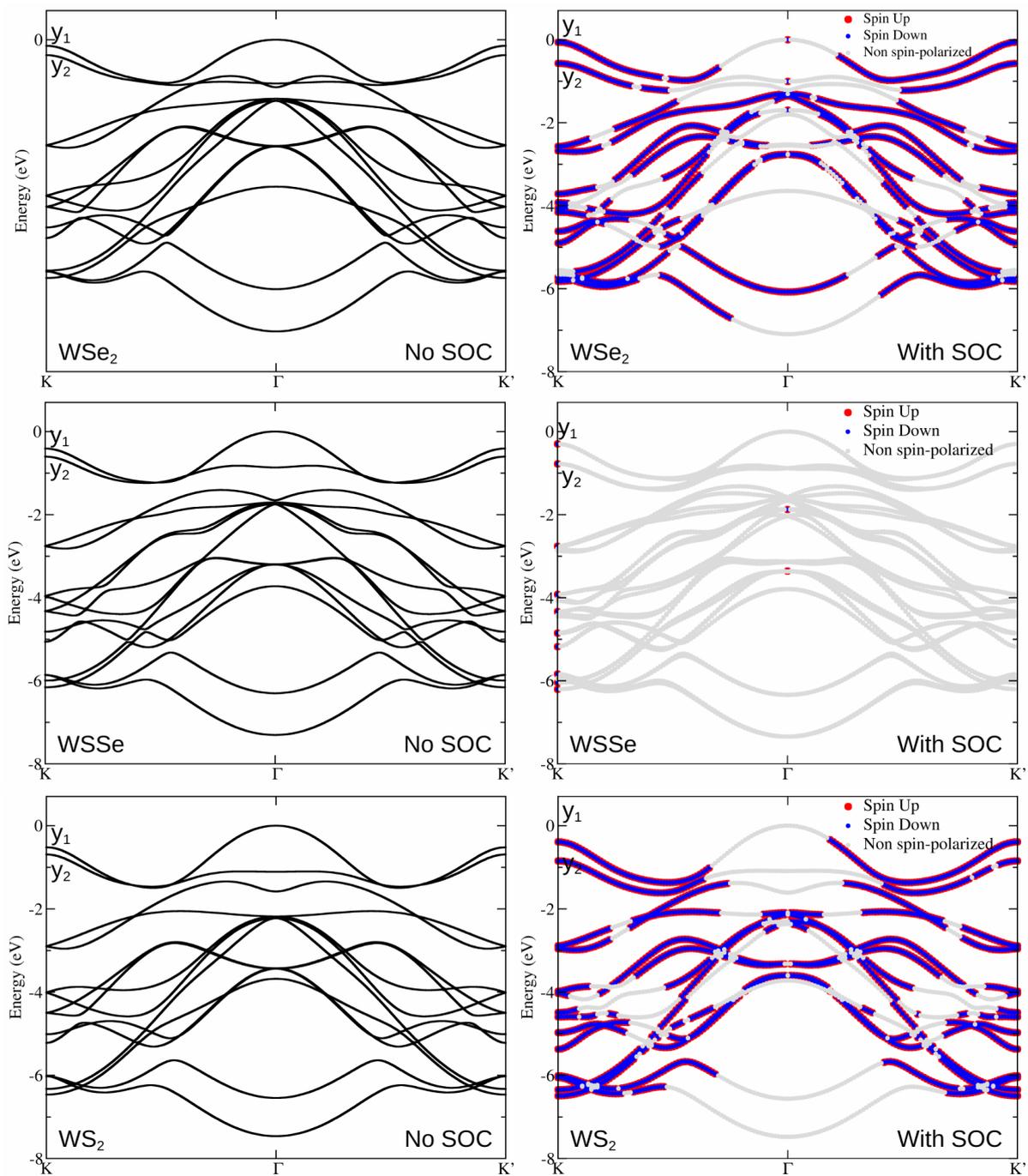

**Figure S4:** Comparison of Band structure for $WSe_2$ (top), Janus WSSe (middle) and $WS_2$ (bottom) without (left) and with SOC (right). The criteria to considered that a spin state is polarized is $|S_z| > 0.35$.

Table S1 reports the energy position of the 2 topmost valence bands at the K point, once the energy of the VBM at Γ has been set to zero.

|  | WSe$_2$ No SOC | WSe$_2$ SOC | WS$_2$ No SOC | WS$_2$ SOC | WSSe No SOC | WSSe SOC |
| --- | --- | --- | --- | --- | --- | --- |
| y1 (eV) | -0.148 | -0.059 | -0.520 | -0.391 | -0.410 | -0.298 |
| y2 (eV) | -0.372 | -0.569 | -0.691 | -0.848 | -0.602 | -0.776 |
| Dy (meV) | 224 | 510 | 171 | 457 | 192 | 478 |
| Dy(SOC)-Dy(NoSOC) (meV) | 286 |  | 286 |  | 286 |  |

Table S1: Energy levels at K(K') points of the Brillouin zone for the two highest valence bands, for WSe$_2$, Janus WSSe and WS$_2$, with and without spin-orbit coupling. The VBM at Γ has been set to zero.

Table S2 cell parameters of the 3 bulk 2H compounds

|  | WSe$_2$ | WS$_2$ | WSSe |
| --- | --- | --- | --- |
| a (bohrs) | 6.225 | **5.991** | 6.107 |
| c (bohrs) | 24,132 | **23.663** | 24.132 |

**Table S2: cell parameters of the 3 bulk 2H compounds**